\documentclass[sigconf]{acmart}
\setcopyright{rightsretained}

\makeatletter
\renewcommand{\@copyrightpermission }{* Both authors contributed equally to this research.
  \\ This work was partially supported by the National Science Foundation (NSF) under award CCF-2140154.}
\makeatother

\copyrightyear{2022}
\acmYear{2022}
\setcopyright{rightsretained}
\acmConference[ICCAD '22]{IEEE/ACM International Conference on
Computer-Aided Design}{October 30-November 3, 2022}{San Diego, CA,
USA}
\acmBooktitle{IEEE/ACM International Conference on Computer-Aided
Design (ICCAD '22), October 30-November 3, 2022, San Diego, CA, USA}
\acmDOI{10.1145/3508352.3549356}
\acmISBN{978-1-4503-9217-4/22/10}

\AtBeginDocument{%
  \providecommand\BibTeX{{%
    \normalfont B\kern-0.5em{\scshape i\kern-0.25em b}\kern-0.8em\TeX}}}

\usepackage{amsmath, amsfonts}
\usepackage{algorithmic}
\usepackage{graphicx}
\usepackage{textcomp}
\usepackage{xcolor}
\usepackage{multirow}

\definecolor{darkgreen}{RGB}{0,120,0}

\DeclareMathOperator*{\argmax}{arg\,max}

\begin{document}

\title{\textsc{Romanus}: Robust Task Offloading in Modular Multi-Sensor Autonomous Driving Systems}



\author{Luke Chen*, Mohanad Odema*, Mohammad Abdullah Al Faruque}

\affiliation{%
  \institution{Department of Electrical Engineering and Computer Science}
  \city{University of California, Irvine}
  \state{California}
  \country{USA}
}%

\begin{abstract}
Due to the high performance and safety requirements of self-driving applications, the complexity of modern autonomous driving systems (ADS) has been growing, instigating the need for more sophisticated hardware which could add to the energy footprint of the ADS platform. Addressing this, edge computing 
is poised to encompass self-driving applications, enabling the compute-intensive autonomy-related tasks to be offloaded for processing at compute-capable edge servers. Nonetheless, the intricate hardware architecture of ADS platforms, in addition to the stringent robustness demands, set forth complications for task offloading which are unique to autonomous driving. Hence, we present \textsc{ROMANUS}, a methodology for robust and efficient task offloading for modular ADS platforms with multi-sensor processing pipelines. Our methodology entails two phases: (\emph{i}) the introduction of efficient offloading points along the execution path of the involved deep learning models, and (\emph{ii}) the implementation of a runtime solution based on Deep Reinforcement Learning to adapt the operating mode according to variations in the perceived road scene complexity, network connectivity, and server load. Experiments on the object detection use case demonstrated that our approach is 14.99\% more energy-efficient than pure local execution while achieving a 77.06\% reduction in risky behavior from a robust-agnostic offloading baseline.
\end{abstract}



\maketitle

\section{Introduction}
Because erroneous or delayed responses in self-driving applications can compromise road safety, equipment, and/or the lives of the passengers themselves, Autonomous Driving Systems (ADS) are required to achieve outstanding performances on core driving tasks, such as perception and localization. Consequently, ADS platforms are designed today to run highly-sophisticated algorithms on intricate hardware architectures to realize such desired levels of performance while being robust to any adverse driving contexts. For that, modern ADS platforms have adopted a multi-modal processing approach for constructing ensemble perspectives of driving scenes using a diverse set of sensory inputs, as in how the Tesla Autopilot systems possess 8 cameras and 12 ultrasonic sensors \cite{tesla}. 

Such a multi-sensor approach leads to the generation of an enormous volume of high-dimensional data that requires tremendous resources for real-time processing, further adding to the power demands of the entire system. Addressing this, a heterogeneous collection of hardware components, as in Application-Specific Integrated Circuits (ASICs) and GPUs, are commonly integrated onto ADS platforms to balance performance demands and power efficiency \cite{lin2018architectural}. Still, hardware advancements are met with growing algorithmic complexity and the requirement for supporting new features, leading the power footprint to remain relatively high. For instance, if we compare two generations of ADS platforms: the Nvidia Drive PX2, which was used by Tesla and Audi Q7 for their autopilot programs \cite{px2tesla, px2audi}, against its successor, the Nvidia Drive AGX Orin \cite{Abuelsamid2020orin}, we find that performance efficiency aside, the baseline power demands increased from 250 W to 800 W, which in theory can have adverse effects on both the thermal comfort of the passengers and the vehicle's driving range \cite{lin2018architectural}.
\vspace{-1.2ex}
\begin{figure}[!h]
\centering
{\includegraphics[,width = 0.46\textwidth]{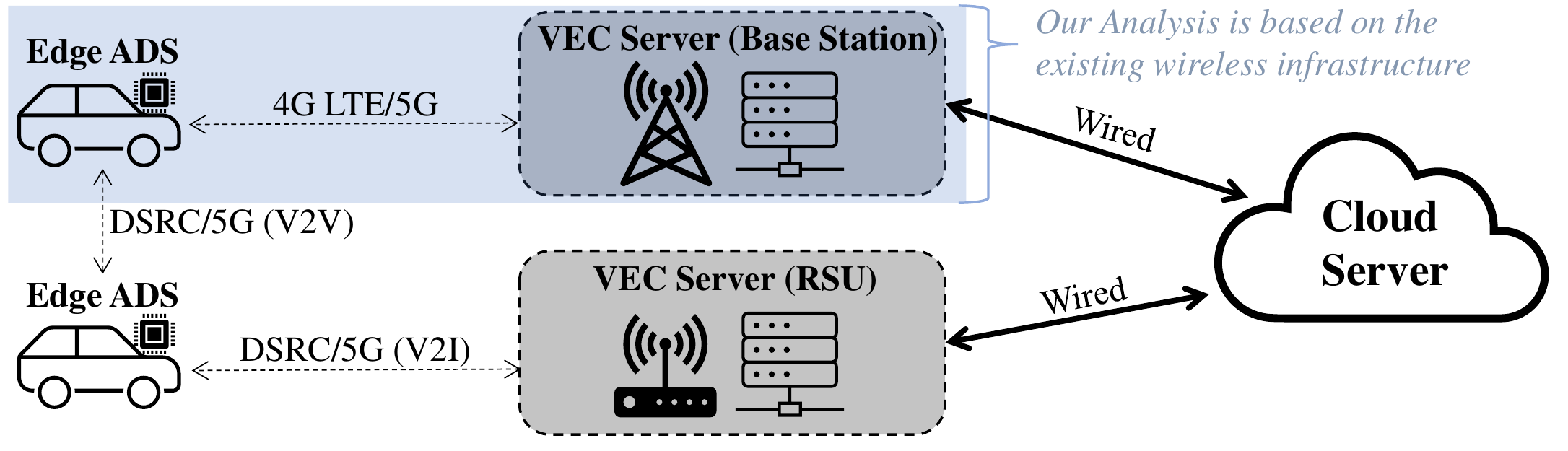}}
\vspace{-1.5ex}
\caption{Vehicular Edge Computing (VEC) Architecture}
\label{fig:hierarchy}
\vspace{-2ex}
\end{figure}

Given how the bulk of processing in ADS modules is largely dominated by deep neural networks (DNNs), compression techniques, e.g., quantization and pruning, have been considered to reduce the modules' complexity, and in turn, their resource requirements  \cite{li2018auto-tuning}. However, experienced performance degradation poses a concern with regards to adopting such techniques' for this class of critical applications.
Alternatively, recent research efforts have targeted exploiting the emerging edge computing paradigm 
for autonomous driving and other vehicular services, in which cloud computing capabilities are brought to the edge of the network through edge servers deployed close to the edge devices, enabling offloading of cumbersome processing burdens to these edge servers for better resource management \cite{liu2019edge}. In the context of vehicular applications, the edge devices are known as connected vehicles (CVs) and the paradigm is further specified as vehicular edge computing (VEC), or vehicular fog computing (VFC) in other cases \cite{baidya2020vehicular}.

Figure \ref{fig:hierarchy} illustrates the hierarchical architecture of VEC and its reliance on the wireless infrastructure, where the higher bandwidths and ultra-low latencies promised by the forthcoming 5G and Dedicated Short Range Communications (DSRC) technologies are to be instrumental in VEC's wide-scale adoption \cite{liu2019edge}. In this regard, VEC computing servers are expected to be deployed either at Road Side Units (RSUs) -- as part of the V2X paradigm -- or cellular base stations, where recent works have proposed to optimize the offloading process to minimize the overall latency and the energy consumed by the ADS \cite{malawade2021sage, baidya2020vehicular}. Still, we find that current approaches in the literature are lacking in the following departments:
 
 \begin{itemize}
    \item The driving context -- crucial to robustness -- is not factored in the offloading decision. Given how a scene's complexity directly correlates with the risk level, and delayed responses could lead to consequences with different levels of severity
    \item Adapting the offloading load according to the runtime conditions while accounting for the underlying ADS composition (e.g., concurrent sensor processing pipelines) is overlooked
\end{itemize}
\vspace{-2ex}

\subsection{Motivational Example}
\begin{figure}[!tbp]
 \centering
{\includegraphics[,width = 0.46\textwidth]{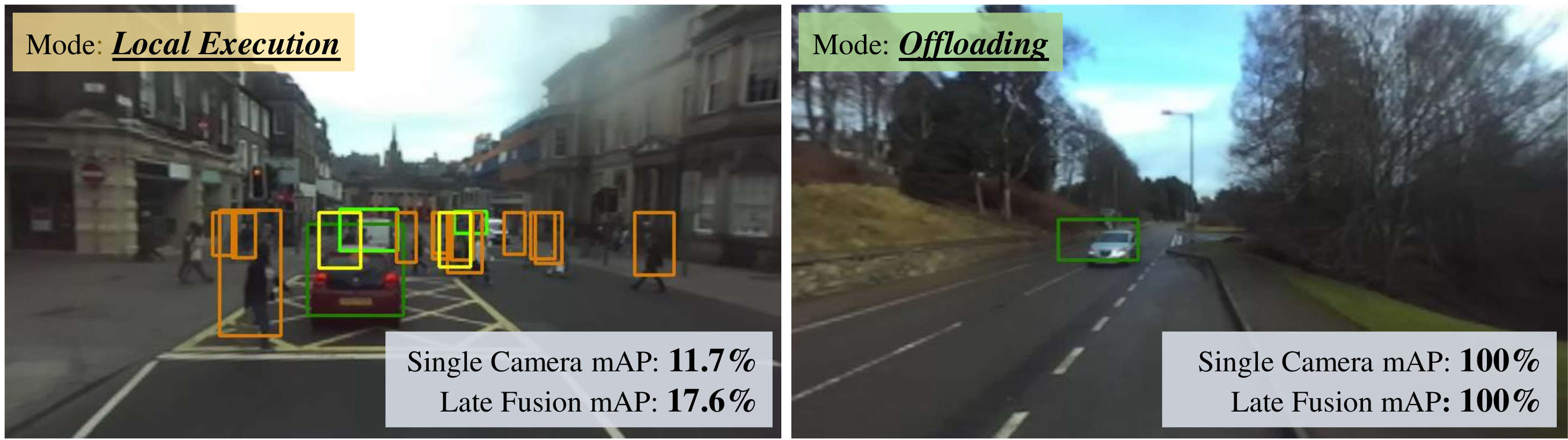}}
\vspace{-2.5ex}
\caption{Two frames of different complexities showing single camera and late fusion mAP scores and the selected operational modes by our learning-based solution. The bounding boxes indicate the ground truths from the dataset.}
\label{fig:motivation}
\vspace{-4.5ex}
\end{figure}

In Figure \ref{fig:motivation}, we show two frames from the Radiate dataset \cite{sheeny2021radiate}, and compare their mean average Precision (mAP) scores on the object detection task. As shown, the left frame instantiates a complex scene with numerous objects of diverse classes, some of them superimposed or obstructed from view leading to relatively low mAP scores. Still, we observe that through fusing the outputs from all sensory pipelines (2 cameras, lidar, and radar), an mAP score of 17.6\% is realized, which surpasses the highest score achieved by a standalone sensory pipeline output -- 11.7\% from the right stereo camera. This alludes to the power of sensor fusion as each sensor can capture its own unique set of features that complement those from other sensors to provide more comprehensive views of the driving scenes. On the flip side, the right frame contains a mere single vehicle that is easily detectable by the standalone camera, achieving a 100\% mAP score. From here, we can contemplate the desired behavior when VEC is supported with regards to tuning the operating mode. Specifically, an ADS experiencing plain driving scenes can opt for \textit{offloading} processing loads from a subset of sensory pipelines for resource efficiency, because although delayed server responses could cause some partial outputs to be absent by the execution deadline, partial fusion of the available local outputs would suffice for this time window due to the relative simplicity of the scene. Contrarily, complicated scenes should have all sensor outputs available for fusion to stimulate robustness, which is only achieved during \textit{local execution} mode as the uncertainty of the wireless networks is avoided. This behavior would be learned by our proposed solution as will be detailed in the following sections.

\subsection{Novel Contributions}

To address the above limitations, we present a methodology for \underline{R}obust Task \underline{O}ffloading in Modular \underline{M}ulti-Sensor \underline{A}uto\underline{n}omo\underline{us} Driving Systems, namely \textsc{Romanus}. From here, we can summarize the main contributions of this paper as follows: 
\begin{itemize}
    \item We present \textsc{Romanus}, a methodology to support efficient and robust offloading for modular ADS platform comprising multiple sensory pipelines with support for sensor fusion. 
    \item As far as our knowledge goes, we are the first to factor the driving context in the offloading decision for the robustness of autonomous driving.
    \item We integrate optimal offloading points within each sensor processing model to realize a dynamic decision space for the runtime operating modes of the ADS.
    \item We implement a Deep Reinforcement Learaning (DRL) based runtime solution that leverages contextual and temporal correlations in the data to optimize the offloading process for latency, energy, and robustness given concurrent pipelines.
    \item Experiments on the object detection use-case using a real-world driving dataset and an industry-grade ADS indicate that our approach is 14.99\% more energy-efficient than local execution while achieving a 77.06\% reduction in risky behavior form a robust-agnostic baseline.
\end{itemize}

\vspace{-1.5ex}

\section{Related Works} \label{sec:related-works}


\textit{\textbf{Mutli-Sensor Perception: }}To maximize information extraction from a driving scene, data is collected from a diverse set of sensors, e.g., cameras, lidar, and radar, to promote perception robustness. Mainly, There are two primary schemes for processing these multi-sensory inputs: \textit{early fusion} \cite{yoo20203d,shahian2019real} and \textit{late fusion} \cite{xu2018pointfusion}. The former combines all sensory features to a single feature at an early point in the ADS pipeline, but is susceptible to sensing noise.
Conversely, the latter offers more resilience at the expense of more redundancy across the sensor pipelines. Recent works \cite{malawade2022hydrafusion, malawade2022ecofusion} have also explored the potential of \textit{hybrid fusion} approaches to leverage the best of both worlds, albeit with added implementation complexities. Here, we concentrate our analysis on the standard \textit{late fusion} approach as it is more challenging and understudied from an offloading perspective.


\textit{\textbf{Vehicular Edge Computing (VEC): }} Numerous research efforts have targeted system-wide resource optimization for VEC through optimal task offloading and scheduling strategies given a variety of servers, vehicles, and tasks \cite{wu2019delay, zhang2016energy}. Typically, such strategies are complemented with runtime solutions that can tune the operation according to variations in the deployment environment, such as the network connectivity conditions \cite{cui2020offloading}. Nonetheless, delayed responses from edge servers are not tolerated in autonomous driving application as the safety of the road, vehicles, and passengers \cite{baidya2020vehicular} can be compromised. Hence, \cite{wang2019auto} proposed a customized communication protocol for a stable and fast offloading of autonomous driving tasks. Even more so, the authors in \cite{malawade2021sage} proposed a fail-safe routine to re-invoke local computation if responses are delayed beyond a certain threshold to account for the uncertainty of wireless links. Such schemes would be even more convoluted when offloading from multiple concurrent pipelines is considered.

\textit{\textbf{DNN Split Computing: }}To identify optimal offloading points within DNN architectures, \cite{neurosurgeon, odema2021lens} analyzed the expected computation and communication costs for each potential offloading layer. For a considerable number of architectures, either direct raw inputs offloading or pure local execution represented the most efficient option. Therefore, works in \cite{eshratifar2019bottlenet, matsubara2019distilled, matsubara2020head} proposed to modify a DNN's structure to include an early optimal offloading layer that shrinks the size of transmissible data, minimizing the costs of both computation and communication. This split-computing concept was applied for end-to-end control in autonomous vehicles \cite{malawade2021sage}, and here, we extend its applicability to multi-sensor modular ADS platforms.


\section{System and Problem Overview} 

\subsection{Autonomous Driving System Composition}

For perspective, we briefly describe the primary modules that compose a state-of-the-art ADS shown in Figure \ref{fig:av-pipeline} as follows:
 
 \begin{figure}[!tbp]
 \centering
{\includegraphics[,width = 0.46\textwidth]{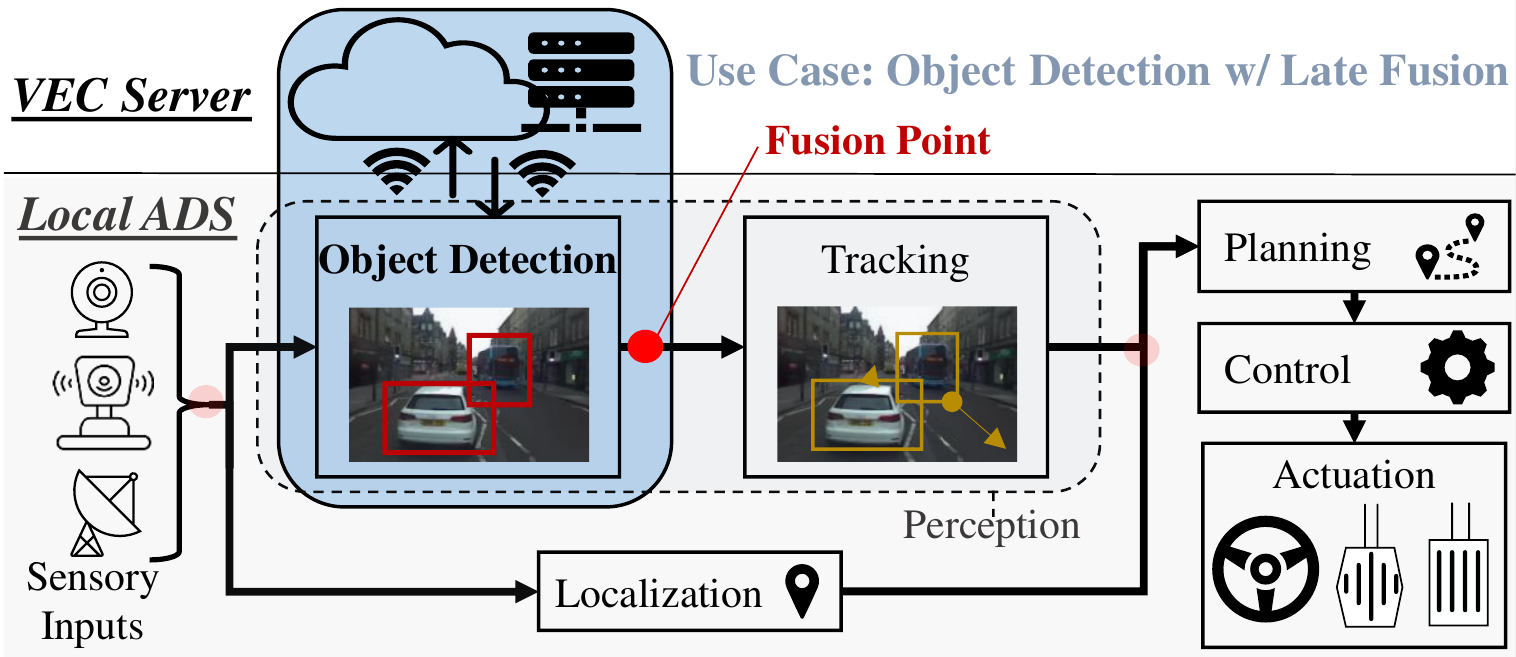}}
\vspace{-1ex}
\caption{Modular industry-grade ADS. The fusion point following the object detection module is this work's use-case.}
\label{fig:av-pipeline}
\vspace{-3ex}
\end{figure}
\textbf{\emph{Perception:}} As the main receptor of the raw sensory data, the perception module is responsible for processing the data over two successive computing blocks. The first is an \textit{object detector} to identify and classify objects of interest, e.g., pedestrians and vehicles, that surround the ego vehicle. A \textit{tracking} module ensues to receive identified objects and associate them with their past movements so as to predict the current movement trajectory.  

\textbf{\emph{Localization:}} Another module taking in the raw inputs is the localization module, whose task is to pinpoint the position of the vehicle at high precision using SLAM/GPS modules.

\textbf{\emph{Planning:}} Outputs from the perception and localization blocks are fused together onto the same 3D co-ordinate space for further processing by the behavioral and motion planning block. From here, a series of sequential path information can be generated from starting position until the endpoint.

\textbf{\emph{Control:} }The final block tasked with mapping the information generated from the planning block onto control instructions for the actuators (e.g., driving wheel, brakes, accelerator). 

\subsection{Problem Formulation} 

In a modular ADS pipeline, the perception block is the dominant entity affecting the end-to-end performance the most \cite{lin2018architectural}, and thus, directing offloading optimizations at this module can lead to substantial efficiency gains across the entire system. Still, sub-optimal operating points can be reached if the following two aspects are not considered properly: (\emph{i}) the nominal safety considerations of the autonomous driving application, and/or (\emph{ii}) the structural composition of the ADS modules themselves. 
For the former, an ADS is required to conclude end-to-end processing under stringent execution time limits to maintain road safety -- a 100 ms deadline at the worst \cite{lin2018architectural, baidya2020vehicular}. Hence, when VEC is supported, expected additional delays due to wireless channel impairments should be considered as part of the overall end-to-end latencies to determine the best offloading decision. Still, additional abrupt delays could threaten the integrity of the self-driving application considering the tightness of the execution windows. 
Whereas for the latter, understanding the underlying structure of an ADS is key to determine the optimal placement of an offloading point that effectively balances the inherent trade-off between communication and computation. For instance, offloading prior to the fusion point can incur a substantial transmission overhead, as opposed to offloading after it which could incur a sizeable computational overhead due to prolonged periods of local processing. 


Formally, a module employing late fusion comprises $N$ processing models $\{f_1, f_2, ..., f_N\}$ for every supported sensor. Thus, for an input vector $X:=x_{1:N}$, the fusion block output can be given by: 
\begin{equation}
    {y} = \mathcal{H}(f_1(x_1), f_2(x_2), ...., f_N(x_N))
    \label{eqn:late_fusion}
\end{equation} 
where $\mathcal{H}$ is the fusion algorithm whose inputs are the $N$ outputs $f_i(x_i) \forall i \in N$. When offloading is supported, the goal is to avoid excessive computational overheads. Thus, each model $f_i$ would incorporate an offloading point to be further defined as: 
\begin{equation}
f_i(x_i) = f_i^T(f_i^H(x_i))
\end{equation}
where $f_i^H$ and $f_i^T$ are the head and tail parts of the $i_{th}$ model placed prior to and after the offloading point, respectively. The former sub-model is to be deployed locally while the latter is to be replicated across the local and edge server platforms. As server responses could peak due to the wireless channel uncertainty, some model outputs may \textit{not be} \textit{available} for fusion given the strict execution deadlines. Thus, we obtain instead partial fusion outputs given by:
\begin{equation}
    {\hat{y}} = \mathcal{H}(\mathcal{I}_1\times f_1(x_1), \mathcal{I}_2\times f_2(x_2), ...., \mathcal{I}_N \times f_N(x_N))
\end{equation} 
where the random variable $\mathcal{I}_i \in \{0,1\}$ indicates whether $f_i(x_i)$ is available for fusion. Naturally, the lesser number of inputs available the more robustness is compromised. Therefore, given $M$ operating offloading modes, the objective is to identify the mode satisfying:
\begin{equation}
    \min\limits_{m\in M} E(X|m), \; s.t. \;\; L(X|m) \leq L_{th}, \;\Delta(y, \hat{y}) \leq C_{th}
    \label{eqn:objective}
\end{equation}
where $E(\cdot)$ and $L(\cdot)$ are the respective end-to-end energy consumption and execution latency for processing the input vector $X$ given processing mode $m$. This formulation is regulated by a latency constraint $L_{th}$ for nominal safety, and a robustness constraint $C_{th}$ based on the difference in prediction quality between $y$ and $\hat{y}$. In the following sections, we present our methodology for solving the optimization objective in (\ref{eqn:objective}), which will entail applying DNN structural optimizations and a runtime learning-based approach. We demonstrate our analysis for the use-case of late fusion following the object detection module as illustrated in Figure \ref{fig:av-pipeline}.

\begin{figure}[!tbp]
{\includegraphics[,width = 0.46\textwidth]{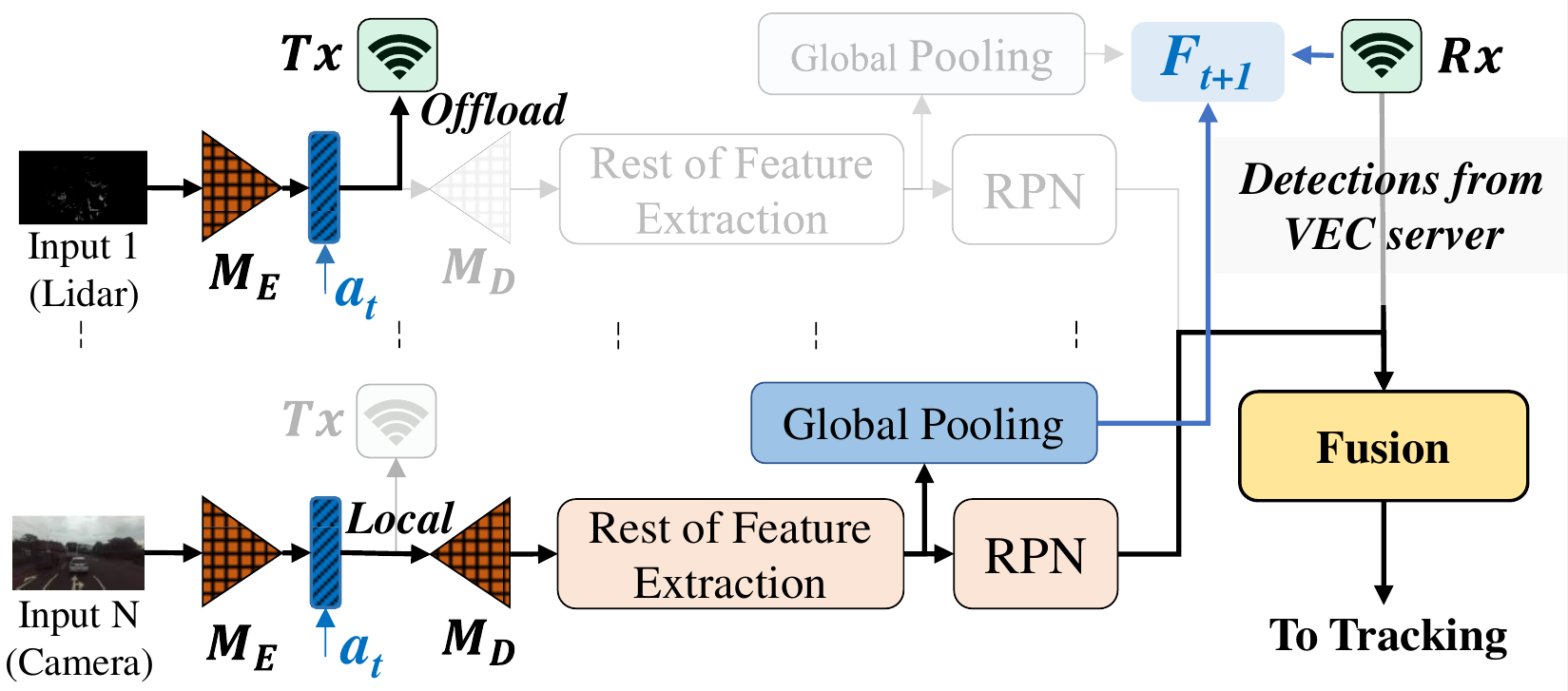}}
\vspace{-2ex}
\caption{Object Detection with Late Fusion and offloading support. Blue blocks/variables are passed to/from the DRL domain in section \ref{sec:DRL}. Transparent blocks are inactive.}
\label{fig:object-detection}
\vspace{-3ex}
\end{figure}

\section{Concurrent Pipelines Design \label{sec:design}}

Figure \ref{fig:object-detection} depicts the proposed processing domain for object detection with our applied modifications given concurrent DNNs and a late fusion scheme detailed as follows. 

\subsection{Object Detection with Late Fusion} 

This scheme entails processing each sensory input separately before aggregating the outputs together through fusion. Specifically, there are two primary computational tasks:


\textit{\textbf{Object Detection:}} For each sensor, an object detection pipeline is implemented to identify and classify objects in a scene. Initially, each model entails a feature extractor based on a Convolutional Neural Network (CNN), e.g., ResNet-18 here \cite{he2016deep}, responsible for abstracting raw sensory data into smaller-sized features for the following detection model, e.g., Faster R-CNN network \cite{ren2015faster}, which consists of a regional proposal network (RPN) to suggest regions of interest where objects may exist, a classification stage to categorize the objects within each proposal, and a final post processing stage to convert classified proposals into bounding box predictions.

\textit{\textbf{Fusion:}} As the outputs from each pipeline are bounding boxes, we can directly fuse them together using Non-Maximum Suppression (NMS) \cite{ren2015faster} to calculate the intersection over union (IoU) and obtain an estimate on the degree of overlapping between each pair of bounding boxes from the overall set of predictions. If an IoU for a pair of boxes exceeds a predetermined threshold, the bounding box with less confidence score is discarded, and this operation repeats until all possible pairing combinations are covered.


\subsection{Implementing DNNs to support Offloading}
To avoid the overhead of offloading raw inputs, we scale the optimal offloading point injection technique in \cite{malawade2021sage, matsubara2020head} to each concurrent pipeline without compromising the overall utility as follows:

\textit{\textbf{Structural Alterations:}} We alter the structure of the feature extractors (ResNet-18 here) within each pipeline to minimize local computation overhead prior to the offloading point and downsize the transmissible data. Specifically, we substitute a considerable portion from the earliest parts of a DNN with an encoder-decoder like structure of two functional components: (\emph{i}) an encoder, $\mathcal{M}_E$, which offers an efficient offloading option at its output through shrinking the input data into a lower-dimensional representation that retains the most relevant of features, i.e., small output sizes translate to low communication overheads., and (\emph{ii}) a decoder, $\mathcal{M}_D$, to receive outputs from the encoder and cast them back to higher dimensional representations of dimensions compatible with the remainder of the network. Here, we replace the first two residual blocks from each ResNet-18 with an encoder-decoder structure. 

\begin{figure}[!tbp]
\centering
{\includegraphics[,width = 0.33\textwidth]{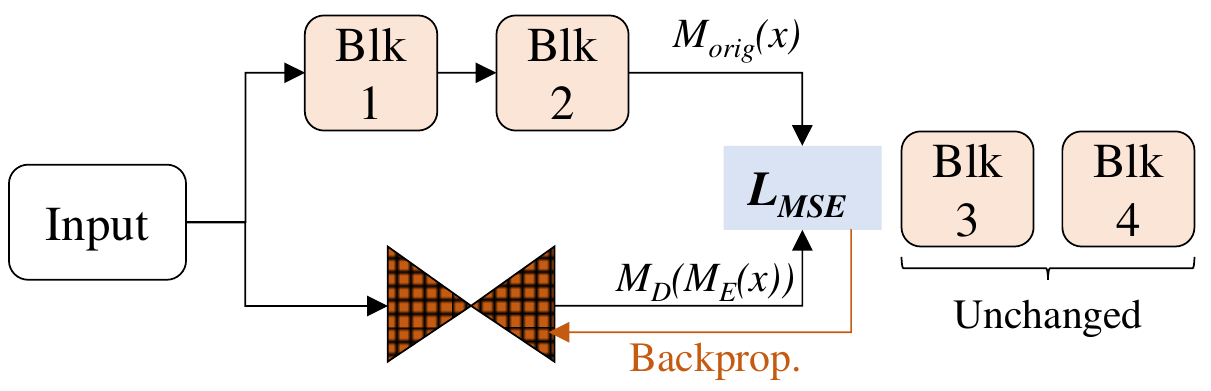}}
\vspace{-2ex}
\caption{Our ResNet-18 feature extractors undergo Knowledge Distillation to train $\mathcal{M}_E\cdot\mathcal{M}_D$ using the first 2 blocks}
\label{fig:KD}
\vspace{-3.5ex}
\end{figure}

\textit{\textbf{Knowledge Distillation:}} Next, modified architectures need to be retrained to maintain utility. We apply knowledge distillation \cite{matsubara2020head} to train $\mathcal{M}_E$ and $\mathcal{M}_D$ through minimizing a loss function, e.g., mean squared error ($L_{MSE}$), between $\mathcal{M}_D$ outputs and those from the original parts, $\mathcal{M}_{orig}$. Figure \ref{fig:KD} illustrates this for our ResNet-18 with the loss component for a single input $x$ given by:   
\begin{equation}
    L_{SE} = || \mathcal{M}_{orig}(x) - \mathcal{M}_D (\mathcal{M}_E(x)) ||_2^{2} \label{eqn:kd}
\end{equation}

Hence, unaltered DNN components can retain their weight values with only the parameters of the new structure trained to produce the same output values as the originals.

\textit{\textbf{Deployment for Inference:}} After retraining, the modified architectures would be deployed for each concurrent detection pipeline on the local ADS. Furthermore, each $\mathcal{M}_D$ and its succeeding blocks would be replicated across the VEC servers to enable online autonomous driving services. Thus during runtime, servers can receive outputs from $\mathcal{M}_E$ components, process them, and return predictions, e.g., bounding boxes coordinates, to the ADS platforms. On the ADS, available local and received predictions are fused to provide the final outputs for the following blocks.

\section{Reinforcement Learning Control} \label{sec:DRL}

VEC operation is reliant on the surrounding conditions with regards to the wireless channel state and the server load. Hence, we propose a learning-based approach based on deep reinforcement learning (DRL) to adapt the mode of operation so as to maximize performance efficiency while maintaining robustness -- which we account for in the offloading decision through leveraging the abstract feature representations already computed within the processing pipelines.

\subsection{Hierarchical Agent}
\begin{figure}[!tbp]
{\includegraphics[,width = 0.49\textwidth]{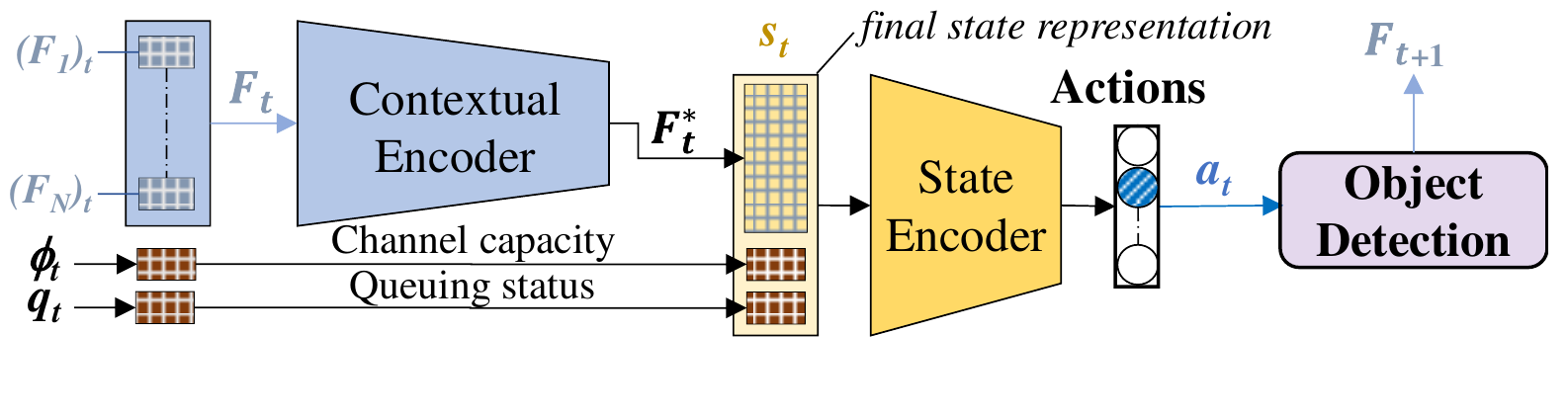}}
\vspace{-6ex}
\caption{Our hierarchical agent for runtime mode selection}
\label{fig:DRL}
\vspace{-3.5ex}
\end{figure}

As shown in Figure \ref{fig:DRL}, our DRL solution constitutes a hierarchical agent whose main components are as follows:

\subsubsection{Contextual Encoder} In order to estimate the complexity of the corresponding scene, we leverage the computed feature set, $\mathcal{F}_{t}$, at time window $t$ from the main sensor processing pipelines to guide the decision for the following window $t+1$, given as $\mathcal{F}_{t} = \{(\mathcal{F}_1)_{t}, (\mathcal{F}_2)_{t}, .., (\mathcal{F}_N)_{t}$\}. The rationale behind using the feature set of the preceding time window is twofold: (\emph{i}) features do not need to be computed from scratch as they have already been generated within the primary processing pipelines (see the global pooling blocks in Figure \ref{fig:object-detection}), and, (\emph{ii}) the small window size for autonomous driving ($\leq$ 100 ms) means that successive frames share similar driving contexts due to the high spatio-temporal scenic correlations.

Given how $\mathcal{F}_{t}$ can outweigh other DRL inputs due to its relatively larger size, $\mathcal{F}_{t}$ needs to be initially encoded into a further lower-dimensional representation. Hence, $\mathcal{F}_{t}$ is processed through a \emph{contextual encoder} comprising a sequence of fully-connected layers to obtain the final abstraction $\mathcal{F}^{*}_{t}$. In our experiments, $\mathcal{F}^{*}_{t}$ was of 256$\times$ smaller in size than $\mathcal{F}_{t}$.


\subsubsection{State Encoder} The next component is the state encoder whose input is the final state representation $s_{t} = \{\mathcal{F}^{*}_{t}, \phi_{t}, q_t\}$ formed from aggregating the contextual encoder outputs, $\mathcal{F}^{*}_{t}$, the channel capacity $\phi_{t}$, and server queuing delays $q_t$. Practically, the latter two metrics can be estimated by probing the edge server. 

\subsubsection{Action Space} Represented by the final fully-connected layer in the \emph{state encoder}, the action space covers the set of all possible modes of operation that can be selected by the DRL at runtime. We define it as $A = \{offload_{0}, offload_{1}, offload_{2}, ... offload_{N-1}\}$, where an action $offload_{i}$ is for choosing the offloading option for $i$ sensory pipelines, with $offload_{0}$ being pure local execution. In the case that the same DNN structure is shared across all pipelines, only the number of offloading pipelines matter. We do not consider $offload_{N}$ as a viable option so that the ADS is always guaranteed a new output every time window since at least one pipeline is always processed locally. This way, even under a worst-case scenario when tasks from $N-1$ pipelines are offloaded and results are not received within the time limit, the vehicle can still operate in a safe manner. In practice, we merely need a subset of actions $A^* \subseteq A$, with \{$offload_{0}, offload_{N-1}\} \subseteq A^*$, where $A^*$ can contain the actions that exhibit notable variability in performance. At runtime, action vector, $a_{t}$, is mapped onto the control of each processing pipeline.

\subsection{DRL Environment}

We detail the emulated DRL training environment for learning a policy $\pi$ that makes offloading decisions based on the current state. 

\subsubsection{Training and Reward}

Reinforcement learning approaches rely on having a Q function to provide value estimates for each state-action pair so as to select the optimal action $\hat{a}$ = $\argmax_{a \in A} Q_{\pi}(\hat{s}, a)$ for each $\hat{s}$ under a learnt policy $\pi$. However, estimating state-action pair values in continuous state spaces is challenging, and DRL offers to approximate $Q_{\pi}$ by a policy network trained to maximize a reward. With no loss in generality, our DRL employs a Double Deep Q-Network \cite{van2016deep} with a compounded reward function $\mathcal{R}$ as:
\begin{equation}
    \mathcal{R} = 
    \begin{cases}
        \mathcal{A}, \;\;\;\;\;\;\;\;\;\;\; \text{if mAP($y$) $<$ mAP$_{th}$} \\
        \mathcal{B}, \;\;\;\;\;\;\;\;\;\;\; otherwise
    \end{cases}
\end{equation}
which evaluates to different functions based on a measure of robustness, which we associate here with the degree of uncertainty in the final fused predictions $y$ in (\ref{eqn:late_fusion}), determined by the mean Average Precision (mAP) scores for object detectors as in \cite{rahman2021per}. In brief, our goal is for the agent to realize a policy that deters from offloading actions when prediction confidence is low, which we achieve here through leveraging the contextual information in $\mathcal{F}_{t-1}$ to assess the scene's complexity, and make offloading decisions accordingly with the goal of minimizing prediction uncertainty. Thus, if $mAP_{th}$ is not met, $\mathcal{R}$ evaluates to $\mathcal{A}$ defined as:
\begin{equation}
    \mathcal{A} = 
    \begin{cases}
        0, \;\;\;\;\;\;\;\;\;\; \text{if $\hat{a} == offload_0$} \\
        \frac{P}{N-i},\;\;\;\;\; \text{if $\hat{a} == offload_i$}
    \end{cases}
    s.t., i\neq 0, i<N
\end{equation}
for penalizing the agent whenever an offloading action is selected, with the penalty value being proportionate to the number of offloading pipelines, $i$, out of $N$ total, with a maximum \textit{negative} penalty of $P$. Recall that $offload_N \not\in A$ as one pipeline always executes locally to ensure  at least one output is available irrespective of the wireless network conditions. On the flip side, when mAP$_{th}$ is satisfied, $\mathcal{R}$ evaluates to $\mathcal{B}$ as follows:
\begin{equation}
    \mathcal{B} = 
    \begin{cases}
        P, \;\;\;\;\;\;\;\;\;\;\; \text{if $L(X|\hat{a}) > L_{th}$} \\
        \mathcal{C}, \;\;\;\;\;\;\;\;\;\;\; otherwise 
    \end{cases}
\end{equation}
which penalizes the agent by $P$ when its selected action $\hat{a}$ causes the overall execution latency for inputs $X$, $L(X|\hat{a})$, to exceed the critical execution latency constraint, $L_{th}$. In other words, this means that the agent is penalized when not all partial outputs are available in time for late fusion.
In reality, state-of-the-art ADS platforms are designed to meet the application latency demands, and hence, we set the value of $L_{th}$ to that of \textit{local execution}.  Contrarily, when $L_{th}$ is satisfied, $\mathcal{R}$ finally evaluates to $\mathcal{C}$ given by:
\begin{equation*}
    \mathcal{C} = 
    \begin{cases}
        0, \;\;\text{if $E(X|\hat{a}) == min(E(X|a)|L(X|a) \leq L_{th})$ } \\
        P, \;\; otherwise
    \end{cases}
\end{equation*}
\begin{equation}
    \forall a \in A^*, A^*\subseteq A  
\end{equation}
penalizing the agent by $P$ if the energy consumption footprint $E(X|\hat{a})$ from selecting action $\hat{a}$ is not the minimal from amongst those of all viable actions $a\in A^*$ that are projected to meet $L_{th}$. 



\subsubsection{Latency and Energy Estimation \label{subsec:latency_energy}}

In order to evaluate $\mathcal{R}$ for each selected $\hat{a}$, the end-to-end estimates for energy and latency can be approximated every time window as follows:
\begin{align}
    L &= L_{local} + L_{Tx} + L_{server} + L_{Rx} \\
    E &= E_{local} + E_{Tx} + E_{idle} + E_{Rx}
\end{align}
where the latency $L$ can be broken down into the respective local, transmission, server, and receiving latencies. Similarly, energy consumption constitutes the same components except for incorporating idling energy as we are only concerned about the ADS energy footprint. From here, the local components are given by:
\begin{align}
    L_{local} = 
        N \times L_{\mathcal{M}_E} + (N-i) \times L_{tail}\mid \text{$\hat{a} == offload_{i}$}
\end{align}
in which $L_{\mathcal{M}_E}$ and $L_{tail}$ are the respective latencies for executing the encoder $\mathcal{M}_E$ and the remaining tail parts of the model, respectively. When the selected action is to offload processing from $i$ processing pipelines (i.e. $\hat{a}$ $== offload_{i}$), the total local execution latency accounts for processing across the $N$ encoders and the $N-i$ tail models. This additive form represents the most direct approach for modeling local execution. However in reality, the concurrency of pipelines can speed up local execution depending on the available hardware resources at the expense of a larger power consumption footprint, $P_{local}$. We approximate this trade-off through considering energy for performance evaluation, defining $E_{local}$ as:
\begin{equation}
   {E}_{local} = L_{local} \times P_{local} \label{eqn:Elocal}
\end{equation}

\subsubsection{Channel Estimation}To estimate the communication overheads, we first fit a Rayleigh distribution curve with scale $\sigma$ to throughput traces $\Phi$ collected from the real-world for different wireless technologies, i.e., $\Phi \sim \text{Rayleigh}(\sigma)$. Then, we use the constructed distribution to sample independent and identically distributed (i.i.d.) random variables as the channel capacity $\phi$ to be used for the training and evaluation processes of the DRL agent where data transmission parameters can be evaluated as:
\begin{equation}
    L_{Tx} =
        \frac{i \times {b}}{\phi}\mid\text{$\hat{a} == offload_i$};\ E_{Tx} = L_{Tx} \times P_{Tx} \label{eqn:tx_latency}
\end{equation}
where $b$ is the transmissible data size from one sensory pipeline while $P_{Tx}$ is the transmission power incurred by the ADS. Similarly, the formulation for the receiving parameters, $L_{Rx}$ and $E_{Rx}$, can be provided given corresponding estimates for channel capacity and data sizes in the downlink.

\subsubsection{Server Queuing\label{eqn:serverQing}}
Lastly, we model the server latency $L_{server}$ using queuing delays where we have:
\begin{equation}
    q_{c} = \frac{(1-\rho)(\rho)^{c}}{1-\rho^{C+1}}
\end{equation}
representing the probability that the offloaded task would encounter $c$ other tasks before it in the server's processing queue, with 
$\rho$ being the average server load, and $C$ being the queue size. From here, we are able to generate a probability density function (pdf) for values within 0-C from which we can sample queuing positions, and consequently approximate $L_{server}$.

\section{Experiments and Results}

\subsection{Experimental Setup}

\subsubsection{Dataset}
We use the RADIATE multimodal perception dataset \cite{sheeny2021radiate} for its diverse driving scenarios and adverse weather conditions such as snow, fog, and rain. The dataset covers 8 object classes with annotations from a Navtech CTS350-X radar, a Velodyne HDL-32e LiDAR, and a ZED stereo camera. The variety of scenes provides a varying degree of difficulty for ADS and enables the robustness assessment. For instance, cameras obstructed by snow offer poor visibility indicating higher difficulty that can cause sub-optimal object detection performance. Here, we implemented 4 object detection DNN pipelines: 2 stereo cameras, radar, and lidar. All inputs are mapped onto the forward-facing perspective for late fusion.

\subsubsection{Training and Metrics}
As was mentioned in Section \ref{sec:design}, the original processing pipelines for each sensing modality comprise a ResNet-18 followed by a Faster R-CNN. These models were trained using a batch size of 1, learning rate of 0.005, and the multi-task loss function in \cite{ren2015faster} which combines both classification and box regression losses. For the NMS fusion, we use a fusion IoU threshold of 0.4. We employ mAP as our evaluation metric with boxes IoU $\geq$ 0.5 since it is widely adopted for object detection tasks \cite{everingham2010pascal} where the average precision is estimated using the precision and recall values. More details about evaluating these values are in \cite{everingham2010pascal,ren2015faster}. 

\subsubsection{Hardware and Performance Evaluation} We use the industry-grade Nvidia Drive PX2 Autochauffer as our ADS hardware. The concurrent DNN models are compiled using the TensorRT library becoming inference engines. The local execution power $P_{local}$ is estimated as the difference in the ADS power measurements when processing and idling. For the transmission power $P_{Tx}$, we follow \cite{malawade2021sage} and evaluate it using the data transfer power models in \cite{close}. 

\subsubsection{Encoder-Decoder Structure} The input frame's resolution for each of the sensory pipelines is 672 $\times$ 376 ($\approx$ 740.25 kB). The encoder, $\mathcal{M}_E$, comprises 3 layers (2 convolutional and 1 pooling), each with a stride of 2 with only 3 channels at the output. Therefore, when the outputs from $\mathcal{M}_E$ are quantized to 8 bits for offloading \cite{malawade2021sage}, the transmissible data size $b$ in equation \ref{eqn:tx_latency} becomes $\approx$ 11.57 kB (64$\times$ less than the input's). The decoder $\mathcal{M}_D$ mimics the structure presented in \cite{matsubara2020head} to have its output of the same dimensions as that from the original second ResNet-18 block.

\subsubsection{DRL Settings} For safety, we always execute the radar pipeline locally \cite{liu2017computer} and define $A^* = \{offload_0, offload_2, offload_3\}$. We set $P=-2$, $C=4000$, $\rho = 0.9$, and mAP$_{th}$ = 0.68 unless otherwise stated. We set $L_{th}=68.12$ based on pure local execution latency. 

\begin{table}[ht]
    \centering
    \caption{Loss and mAP (\%) before (orig) and after (dist) integrating $\mathcal{M}_E\cdot \mathcal{M}_D$ across various late fusion combinations.}
    \vspace{-2ex}
    \begin{tabular}{l | c c | c c}
    \hline
    Sensor & Loss (orig) & Loss (dist) & mAP (orig) & mAP (dist) \\
    \hline
    2 Cameras & 0.15 & 0.17 & 67.14 & 67.14 \\
    Radar+Lidar& 0.10 & 0.11 & 67.14 & 67.14\\
    Full Fusion & 0.13 & 0.15 & 71.24 & 70.38 \\ 
    \hline
    \end{tabular}
    \label{tab:map}
\end{table}
\vspace{-3ex}

\subsection{Object Detection and Performance}
We first assess how the inclusion of $\mathcal{M_E}$ and $\mathcal{M_D}$ impacts the loss and prediction accuracy of object detection. Table \ref{tab:map} shows the changes in these metrics across different late fusion combinations on the Radiate evaluation dataset. As seen, full sensor fusion has the best performance in mAP, asserting how prediction robustness relates to the number of fused outputs. It is also observed that the new DNN structures maintain the same level of performance as their original counterparts, with the highest degradation in mAP from 71.24\% to 70.38\% experienced by the full fusion case, but still offering a better score than that of the simpler sensor combinations.

\begin{table}[ht]
    \centering
    \caption{Hardware Measurements on the Nvidia Drive PX2}
    \vspace{-2ex}
    \begin{tabular}{l c c c c}
    \hline
    DNN & $L_{local}$ (ms) & $E_{local}$ (J) & Memory (MB) \\
    \hline
    Encoder  & 3.78 & 0.03 & 0.025\\
    1 pipeline & 17.03 & 0.12 & 80.3\\
    4 pipelines & 68.12 & 0.48 & 321.2 \\
    DRL Agent & 0.66 & 0.005 & 5.4\\

    \hline
    \end{tabular}
    \label{tab:px2}
\end{table}

Table \ref{tab:px2} displays the processing overheads for different DNN components deployed on the PX2 hardware. The encoder $\mathcal{M_E}$ and DRL agent take 3.78 and 0.66 ms, respectively, emphasizing how the decision $a_t$ is obtained before the generation of any transmissible outputs. Moreover, the execution latency for 4 pipelines on the PX2 can add up to 68.12 ms given the same power $P_{local}$. 



\subsection{Channel Capacity and Queuing Analysis}

\begin{figure}[!tbp]
\centering
{\includegraphics[,width = 0.46\textwidth]{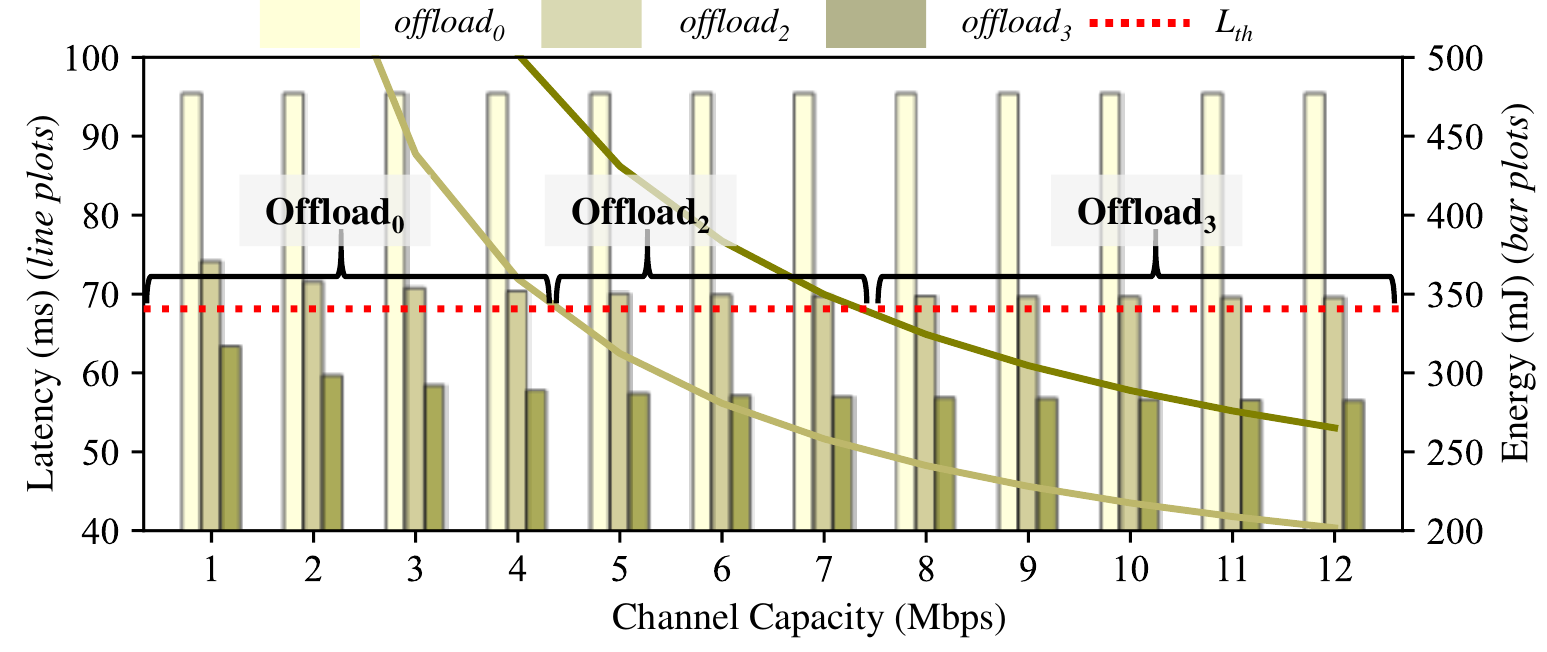}}
\vspace{-2ex}
\caption{Variation of Latency and Energy Analysis w.r.t. Channel Capacity. Energy as bar charts, Latency as plot lines}
\label{fig:channel_sweep}
\vspace{0.5ex}
\end{figure}
\begin{figure}[!tbp]
\centering
{\includegraphics[,width = 0.46\textwidth]{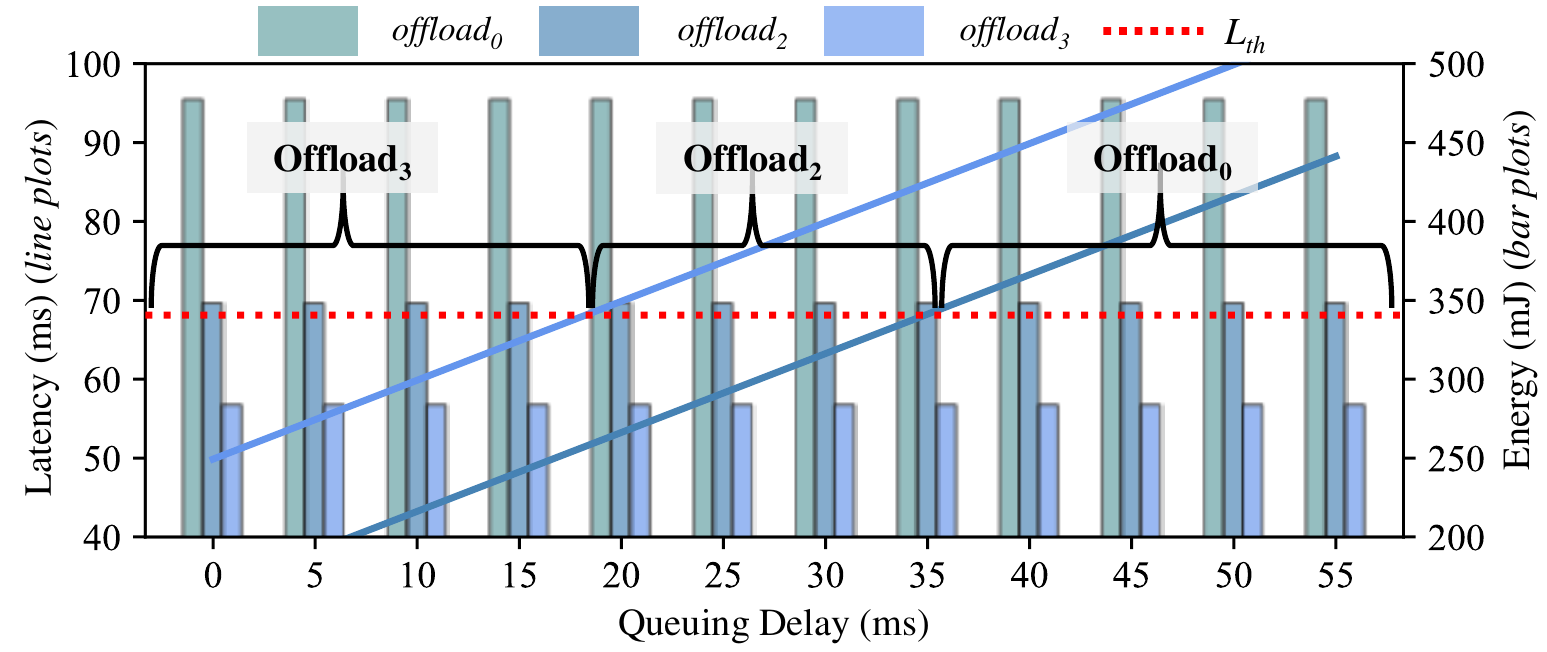}}
\vspace{-2ex}
\caption{Variation of Latency and Energy Analysis w.r.t. Queuing Delay. Energy as bar charts, Latency as plot lines}
\label{fig:queue_sweep}
\vspace{-3.5ex}
\end{figure}

In this experiment, we analyze the influence of the experienced channel capacity, $\phi$, and queuing delay, $q_t$, on the optimal action choice when optimizing for energy consumption under the latency constraint $L_{th}$. To elaborate, we illustrate in Figure \ref{fig:channel_sweep} parametric sweeps with respect to $\phi$ given $q_t=$ 15 ms. As shown, offloading options are consistently more energy efficient than the pure local option (\textit{offload$_0$}), but the $L_{th}$ constraint dictates which action should be chosen considering how poor values of $\phi$ could disqualify some offloading choices. When $\phi>4$ Mbps, the latency overhead for \textit{offload$_2$} does not exceed $L_{th}$ making it the optimal action until $\phi>7$ Mbps, at which the most energy-efficient option, \textit{offload$_3$}, becomes valid. Similarly, this analysis is repeated in Figure \ref{fig:queue_sweep} when sweeping across $q_t$ given $\phi=8$ Mbps. Naturally, the latency overhead is linearly proportional to $q_t$ under fixed network conditions, demonstrating the influence of server load over the optimal offloading decision. From here, the key takeaway is that based on the wireless infrastructure and VEC server capabilities, the maximum number of concurrent offloading pipelines that meet $L_{th}$ can be determined, and used accordingly to construct the decision space.

\begin{table}[ht]
    \centering
    \caption{Robustness analysis at $\rho$=0.97}
    \vspace{-2ex}
    \begin{tabular}{l c c c c }
    \hline
    \multirow{2}{*}{Policy} & \multirow{2}{*}{mAP$_{th}$} & \multicolumn{3}{c}{Average mAP (\%)} \\ \cline{3-5} & &  $offload_0$ & $offload_2$ & $offload_3$ \\ \hline
    R-agnostic & N/A & 64.93 & 64.50 & 64.85 \\ \hline
    \multirow{3}{*}{DRL} &
    0.50 & 60.39 & 67.69 & 68.58 \\ \ &
    0.68 & 60.68 & 70.93 & 72.55 \\  &
    0.98 & 61.82 & 72.49 & 73.14 \\ \hline
    \multirow{3}{*}{Oracle} &
    0.50 & 49.58 & 85.34 & 85.26 \\ \ &
    0.68 & 50.58 & 93.68 & 93.62 \\ \ &
    0.98 & 54.84 & 99.51 & 99.49 \\ \hline
    \end{tabular}
    \label{tab:robustness_mAP}
\vspace{-3ex}
\end{table}

\subsection{Robustness Analysis}
To analyze the DRLs capacity to make robust decisions, we define 2 baseline policies for comparison: (\emph{i}) a robustness-agnostic or \textit{R-agnostic} policy that is aware of $\phi$ and $q_t$ to optimize for energy so long as $L_{th}$ is satisfied, and (\emph{ii}) an \textit{Oracle} resembling an optimal strategy which in addition to the information available to the \textit{R-agnostic} policy, also possesses the true per-frame mAP estimate apriori, and consequently, the optimal sequence of decisions satisfying the robustness constraint mAP$_{th}$. All of the mentioned strategies are more energy-efficient than pure local execution, and the mAP$_{th}$ values are set in the experiments to 0.5, 0.68, and 0.98, estimated based on the cumulative distribution of the evaluation dataset such that 30\%, 50\%, and 70\% of the evaluation mAP scores fall under the corresponding thresholds.

\begin{figure}[!tbp]
\centering
{\includegraphics[,width = 0.46\textwidth]{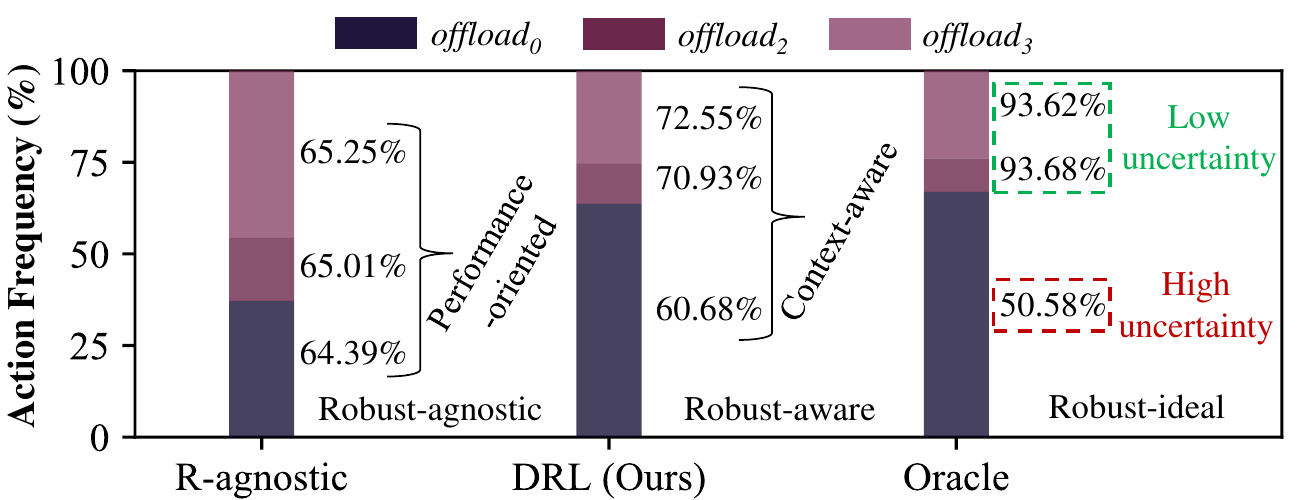}}
\vspace{-2ex}
\caption{Action selection frequencies (\%) breakdown across the 3 policies at mAP$_{th}$=0.68 and $\rho$=0.97. Numbers next to the bars indicate the average experienced mAP (AMAP) for the evaluation dataset inputs mapped to each action.} 
\label{fig:action_map}
\vspace{-3.5ex}
\end{figure}

To evaluate robustness across each policy, we compute the average experienced mAP per action (AMAP) given the action selection frequencies. Mainly, a robust behavior cause frames of high uncertainty (mAP $\leq$ mAP$_{th}$) to be processed locally, implying how the AMAP experienced locally should be low compared to those from the offloading actions. We illustrate this concept in Figure \ref{fig:action_map} across the 3 policies for mAP$_{th}$ = 0.68 and $\rho$ = 0.97. As seen, the \textit{R-agnostic} policy only considers performance efficiency for its action selection, and subsequently, its AMAP across \textit{offload$_0$}, \textit{offload$_2$}, and \textit{offload$_3$}, are equivalent with values of 64.39\%, 65.01\%, and 65.25\%, respectively. Conversely, the $Oracle$ resembles the ideal embodiment of robustness, assigning high uncertainty frames to \textit{offload$_0$}, despite performance gains from offloading. In contrast to the \textit{R-agnostic} policy, \textit{66.94\%} of the \textit{Oracle} policy decisions are \textit{offload$_0$} with an AMAP of 50.58\%, and an AMAP as high as \textit{93.68\%} for the remaining offloading decisions. From here, our proposed \emph{DRL} approach strives to learn the \textit{Oracle}'s behavior, through the observed action selection breakdown, with 63.65\% of actions belonging to \textit{offload$_0$}. Moreover, the AMAP for \textit{offload$_2$} and \textit{offload$_3$} are 70.93\% and 72.55\%, respectively, which despite outperforming the \textit{R-agnostic} policy, are far from that of the \textit{Oracle}. This is expected considering the \textit{Oracle} policy is the unrealistic ideal behavior with apriori mAP knowledge. We extend this analysis to other thresholds values in Table \ref{tab:robustness_mAP}, where we observe that as the robustness constraint becomes smaller, the \textit{DRL} exhibits a behavior closer to the \textit{R-agnostic} and farther from the \textit{Oracle} and vice versa, indicating the \textit{DRL}'s capacity to adapt to various robustness requirements. 
\vspace{-2ex}

\begin{table}[ht]
    \centering
    \caption{Action frequency analysis at mAP$_{th}$ = 0.68}
    \vspace{-2ex}
    \begin{tabular}{l c c c c}
    \hline
    \multirow{2}{*}{Policy} & \multirow{2}{*}{$\rho$} & \multicolumn{3}{c}{Action Frequency (\%)} \\ \cline{3-5}
    & & $offload_0$ & $offload_2$ & $offload_3$ \\ \hline
    \multirow{3}{*}{R-agnostic}
    & 0.90 & 11.17 & 14.73 & 74.10 \\ 
    & 0.97 & 37.18 & 17.17 & 45.65 \\ 
    & 0.99 & 70.34 & 9.25 & 20.41  \\ \hline
    \multirow{3}{*}{DRL}
    & 0.90 & 51.73 & 8.83 & 39.44   \\ 
    & 0.97 & 63.65 & 10.95 & 25.39  \\ 
    & 0.99 & 85.14 & 5.46 & 9.4     \\ \hline
    \multirow{3}{*}{Oracle}
    & 0.90 & 53.19 & 8.00 & 38.82 \\
    & 0.97 & 66.94 & 9.17 & 23.89 \\ 
    & 0.99 & 84.37 & 5.04 & 10.60 \\ \hline
    \end{tabular}
    \label{tab:serverload}
\end{table}

Furthermore, we vary the server load $\rho$ in Table \ref{tab:serverload} and show how the action frequency varies for each policy. As $\rho$ increases, the selection of local processing becomes more frequent across all policies, irrespective of the energy or robustness due to the $L_{th}$ constraint. Such behavior is learned by our $DRL$ solution given how the action selection frequencies closely imitate that of the $Oracle$.
\vspace{-1.5ex}

\begin{table}[ht]
    \centering
    \caption{Energy analysis relative to local at mAP$_{th}$ = 0.98}
    \vspace{-2ex}
    \begin{tabular}{l c c c c}
    \hline
    Metric & Local & R-agnostic & DRL & Oracle \\
    \hline
    Risky Actions (\%) & 0 & 63.37 & \textbf{14.54} & 0 \\
    Robust Actions (\%) & 100 & 36.63 & \textbf{85.46} & 100 \\
    Total Energy (kJ) & 2.916 & \textbf{1.729} & 2.479 & 2.487 \\
    Total Energy Red. (\%) & 0 & \textbf{40.7} & 14.99 & 14.72 \\
    \hline
    \end{tabular}
    \label{tab:energy0.98}
\end{table}

\vspace{-3ex}

\begin{figure}[!tbp]
\centering
{\includegraphics[,width = 0.49\textwidth]{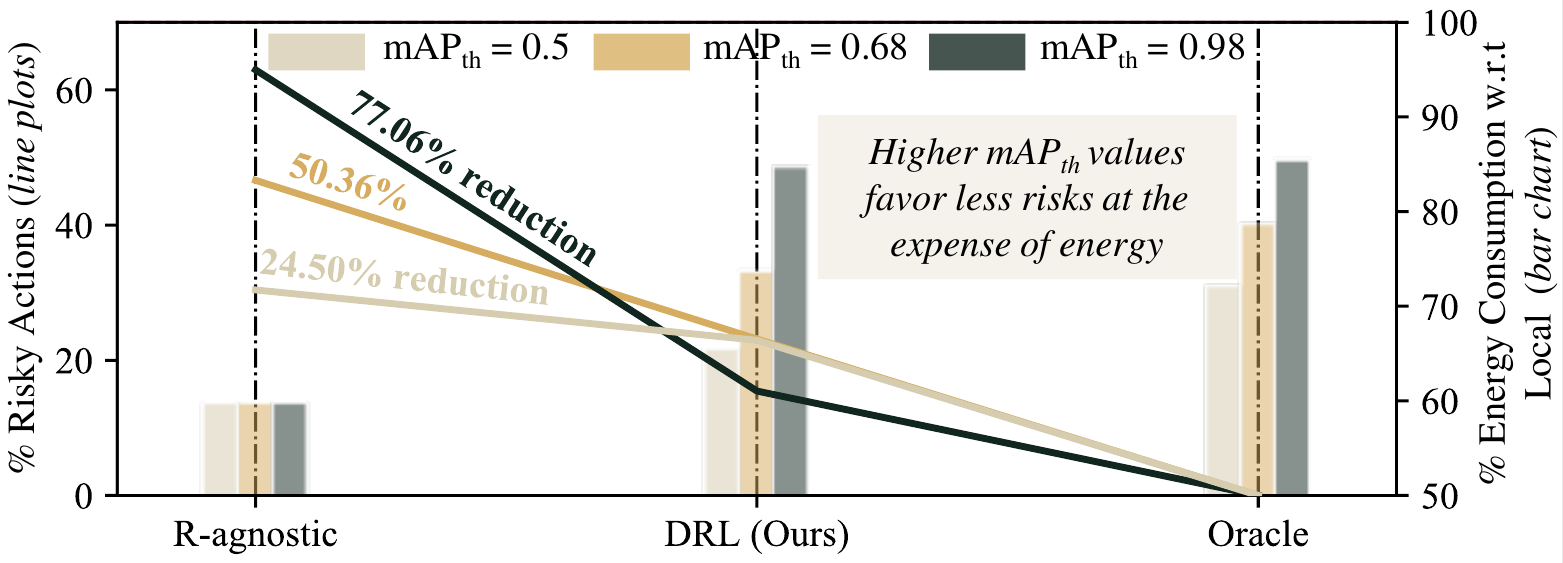}}
\vspace{-5ex}
\caption{Comparing Energy and Risky Actions}
\label{fig:energy_risk}
\vspace{-4ex}
\end{figure}

\subsection{Energy Reduction vs Risky Actions}
We also compare the energy savings relative to the pure local execution, \textit{offload$_0$}, in addition to their risky behaviors. We first define \textit{Risky Actions} as the fraction of offloading actions whose respective mAP scores fall below mAP$_{th}$, and \textit{Robust Actions} as the fraction whose scores exceed the mAP$_{th}$. We compare the performance of each policy in Table \ref{tab:energy0.98}, where although \textit{R-agnostic} offers the highest energy reduction of 40.7\% compared to \textit{DRL}'s 14.99\%, 63.37\% of \textit{R-agnostic}’s energy savings are \textit{Risky Actions}, unlike \textit{DRL} whose \textit{Risky Actions} constitute 14.54\% of the offloading decisions. Through extending this analysis further to entail multiple mAP$_{th}$ values, i.e., a higher threshold means a stricter offloading constraint, we observe in Figure \ref{fig:energy_risk} that the robustness-aware \textit{DRL} at higher mAP$_{th}$ substantially reduces the percentage of risky offloads compared to the \textit{R-agnostic} policy, with reductions of 24.50\% and 77.06\%, at mAP$_{th}=0.50$ and mAP$_{th}=0.98$ respectively.

\section{Conclusion}
In this work, we presented \textsc{Romanus}, a methodology for robust and efficient task offloading for multi-sensor autonomous driving systems (ADS). We first showed how to integrate optimal offloading points along the processing pipelines in a multi-sensor object detection module with late fusion, and then implemented a DRL-based runtime offloading that achieves 14.99\% energy efficiency over pure local execution with up to 77.06\% decrease in risky offloading actions from a robust-agnostic solution. This methodology can be generalized to a variety of sensors and fusion strategies depending on the underlying system structure and how robustness is characterized with regards to the application's primary task, which would form the basis for future research works along this direction.


\bibliographystyle{ACM-Reference-Format}
\bibliography{arxiv-bib}


\end{document}